\documentclass[trackchanges,twocolumn,tighten]{aastex62} 

\usepackage{amsmath}

\usepackage{longtable}
\usepackage{soul}
\usepackage{CJK}
\usepackage{lineno}

\newcommand{\code}[1]{\texttt{#1}}

\newcommand{\mesa}{\code{MESA}}
\newcommand{\MESA}{\mesa}

\newcommand{\stella}{\code{STELLA}}

\newcommand{\Athena}{\code{Athena++}}

\newcommand{\Msun}{M_\odot}
\newcommand{\Rsun}{R_\odot}
\newcommand{\Lsun}{L_\odot}
\newcommand{\vc}{v_c}
\newcommand{\vrms}{v_\mathrm{rms}}

\newcommand{\Teff}{T_\mathrm{eff}}

\newcommand{\rphot}{R_\mathrm{phot}}
\newcommand{\intd}{\mathrm{d}}
\newcommand{\intdt}{\mathrm{d}t}
\newcommand{\intdr}{\mathrm{d}r}
\newcommand{\intdm}{\mathrm{d}m}

\newcommand{\Rib}{R_\mathrm{IB}}
\newcommand{\rshock}{r_\mathrm{sh}}

\newcommand{\y}{y}
\newcommand{\R}{R}
\newcommand{\M}{M}
\newcommand{\E}{E}
\newcommand{\kes}{\kappa_\mathrm{es}}
\newcommand{\vsh}{v_\mathrm{sh}}
\newcommand{\vperp}{v_\perp}
\newcommand{\DR}{\Delta\R}

\newcommand{\tdiff}{t_\mathrm{diff}}

\newcommand{\thalf}{t_{1/2}}
\newcommand{\Dthalf}{\Delta\thalf}
\newcommand{\tcross}{t_\mathrm{cross}}
\newcommand{\trise}{t_\mathrm{rise}}
\newcommand{\tfall}{t_\mathrm{fall}}
\newcommand{\hrho}{H_\rho}
\newcommand{\Mej}{M_\mathrm{ej}}
\newcommand{\Eexp}{E_\mathrm{exp}}
\newcommand{\taus}{\tau_\mathrm{s}}
\newcommand{\tausbo}{\tau_\mathrm{sbo}}
\newcommand{\Tsbo}{T_\mathrm{sh}}
\newcommand{\Lbol}{L_\mathrm{bol}}
\newcommand{\Lchar}{L_\mathrm{char}}
\newcommand{\Lpeak}{L_\mathrm{peak}}

\newcommand{\ltapprox}{\lesssim}

\newcommand{\appropto}{\mathrel{\vcenter{
		\offinterlineskip\halign{\hfil$##$\cr
	\propto\cr\noalign{\kern2pt}\sim\cr\noalign{\kern-2pt}}}}}

\newlength{\apjcolwidth}
\newlength{\figwidth}
\newlength{\doublewide}
\renewcommand{\deleted}[1]{}

\begin{document}
\begin{CJK*}{UTF8}{gbsn}

\title{Shock Breakout in 3-Dimensional Red Supergiant Envelopes} 

\author[0000-0003-1012-3031]{Jared A. Goldberg}
\affiliation{Department of Physics, University of California, Santa Barbara, CA 93106, USA}

\author[0000-0002-2624-3399]{Yan-Fei Jiang(姜燕飞)}
\affiliation{Flatiron Institute, New York, NY, USA}

\author{Lars Bildsten}
\affiliation{Department of Physics, University of California, Santa Barbara, CA 93106, USA}
\affiliation{Kavli Institute for Theoretical Physics, University of California, Santa Barbara, CA 93106, USA}

\correspondingauthor{J. A. Goldberg}
\email{goldberg@physics.ucsb.edu}

\begin{abstract}
Using \texttt{Athena++}, we perform 3D Radiation-Hydrodynamic calculations 
of the radiative breakout of the shock wave in the 
outer envelope of a red supergiant (RSG) which has suffered core collapse and will become a Type IIP supernova. The intrinsically 3D structure of the fully convective RSG envelope yields key differences in the brightness and duration of the shock breakout (SBO) from that predicted in a 1D stellar model. First, the lower-density `halo' of material outside of the traditional photosphere in 3D models leads to a shock breakout at 
lower densities than 1D models. This would prolong the duration of the shock breakout flash at any given location on the surface to $\approx1-2$ hours. However, we find that the 
even larger impact is the intrinsically 3D effect associated with large-scale fluctuations in density that cause the shock to break out at different radii at different times. 
This substantially prolongs the SBO duration to $\approx3-6$ hours and implies a 
diversity of radiative temperatures, as different patches across the stellar surface are at different stages of their radiative breakout and cooling at any given time. 
These predicted durations are in better agreement with existing observations of SBO.
The longer durations lower the predicted luminosities by a factor of $3-10$ ($L_\mathrm{bol}\sim10^{44}\mathrm{erg\ s^{-1}}$), and we derive the new scalings of brightness and duration with explosion energies and stellar properties. These intrinsically 3D properties eliminate the possibility of using observed rise times to measure the stellar radius via light-travel time effects.
\end{abstract}


\section{Introduction}

At the end of a massive ($10\Msun\ltapprox\M\ltapprox25\Msun$) star's life, the 
core collapses and the resulting explosion generates a strong shock, unbinding the hydrogen-rich red supergiant (RSG) envelope.
As the shock nears the outer layers, radiation escapes, leading to a hot 
($T>10^5$K), bright flash, known as the ``shock breakout" (SBO). 
For 1D (spherically-symmetric) models with a well-defined outer radius, 
semi-analytical solutions exist for the shock propagation \citep[e.g.][]{Lasher1979,Matzner1999,Katz2010}, with extensive predictions for the bolometric and optical-UV lightcurves \citep{Nakar2010,Rabinak2011,Sapir2011,Katz2012,Sapir2013,Shussman2016,Sapir2017,Kozyreva2020}. One important prediction of these 1D models is that the observed SBO duration is set by the light-travel time across the stellar surface, $R/c\ltapprox1$hour, which if measured would provide a direct constraint on the stellar radius, $R$. This is a crucial measurement, as it would constrain the ejected mass ($\Mej$) and explosion energy ($\Eexp$) when combined with information from the Type IIP Supernova (SN-IIP) lightcurve \citep{Goldberg2019,Goldberg2020}. 

Observations of SBO from SNe-IIP are presently sparse, with only a few serendipitous detections by NASA's \textit{GALEX} satellite, all of which show durations of $>6$ hours \citep[e.g.][]{Schawinski2008, Gezari2008, Gezari2010, Gezari2015}. This prolonged duration is often attributed to interaction with a dense wind beyond a traditional 1D photosphere
\citep[e.g.][]{Gezari2008,Haynie2021} or by assuming an outer density orders of magnitude lower than traditional 1D models \citep{Schawinski2008}. 
This prolonged SBO is further corroborated by differences between SNe-IIP seen by \textit{Kepler} \citep{Garnavich2016} and \textit{TESS} \citep{Vallely2021,Tinyanont2021} compared to spherically-symmetric SBO models. 
In upcoming years, the data are expected to improve dramatically, as future satellites such as the ULTRASAT instrument \citep{Sagiv2014,Asif2021} are poised to capture hundreds of SN SBO's at high cadence, a number which will grow when combined with data from wide-field X-ray satellites \citep{Bayless2021}. 

SBO in realistic 3D RSG envelopes, which exhibit large-scale coherent plumes spanning large fractions of the stellar surface \citep[e.g.][]{Chiavassa2011a,Goldberg2021}, has not been explored. We show that large-scale, fully convective, intrinsically 3D envelopes yield significant differences in the predicted SBO signal, which has implications for the detectability of these transients, as well as the ability to extract information about the progenitors and explosions from the lightcurves. 
This work is organized as follows:
in \S\ref{sec:SETUP}, we discuss our simulation setup and verification. In \S\ref{sec:FIDUCIAL}, we discuss the shock evolution in a fiducial explosion of one of our models, and in \S\ref{sec:OBSERVABLES}, we discuss the temperature structure of the 3D SBO and properties of the bolometric lightcurves and present our initial energy scalings of the SBO brightness and duration when the 3D surface is taken into account.

\section{Setup and Model Properties}
\label{sec:SETUP}
We use the 3D radiation hydrodynamic (RHD) simulations of RSG envelope models performed by \cite{Goldberg2021} using \Athena\ \citep{Stone2020,Jiang2021} with $X=0.6,~Z=0.02$ (mean molecular weight $\mu=0.645$) and opacities from OPAL \citep{Iglesias1996}. 
Table~\ref{tab:models} summarizes the properties of these models; we treat RSG1L4.5 as our fiducial model. The stellar photospheric radius $\rphot$ is taken where shell-averaged radial profiles of the luminosity $\langle L(r)\rangle$ and radiation temperature $\langle T_r(r)\rangle$ agree, $\langle L(r)\rangle=4\pi r^2\sigma_\mathrm{SB}\langle T_r(r)\rangle^4$ {where $\sigma_\mathrm{SB}$ is the Stefan-Boltzmann constant,} following \citet{Chiavassa2011a}.
\deleted{The RSG1L4.5 simulation is our fiducial model, with a photospheric radius $\rphot=796\Rsun$, inner boundary at $\Rib=400\Rsun$, mass interior to the simulation domain of $12.8\Msun$, and a total stellar mass of $16.4\Msun$.} 
\deleted{The RSG2L4.9 model, (see \S\ref{sec:OBSERVABLES}) has $\rphot=902\Rsun$, $\Rib=300\Rsun$, $M_\mathrm{IB}=10.79\Msun$, and stellar mass $12.9\Msun$.}
We use the RHD scheme presented by \citet{Jiang2021} with the same simulation domain as in \citet{Goldberg2021} for each model. 
This is a spherical polar grid with 120 angles per grid cell for the specific intensities, 128 bins from $\theta=\pi/4-3\pi/4$ and 256 bins from $\phi=0-\pi$ with periodic $\theta/\phi$ boundary 
conditions, covering 70.6\% of the face-on hemisphere, $\Omega=1.41\pi$). For RSG1L4.5 (RSG2L4.9) we use 384 (256) bins in radius, for $r=400-22,400\Rsun (r=300-6,700\Rsun)$, with $\delta r/r=0.01$, and an outflow outer boundary. No initial perturbations are necessary, as fluctuations in the fluid properties are achieved by the full 3D RHD treatment in the convectively unstable envelope in its convective quasi-steady-state. 
For more details about the 3D RSG envelope models, see \citet{Goldberg2021}. 

\begin{table*}
\begin{center}
\begin{tabular}{| c | c | c | c | c | c | c | c | c | }
\hline Progenitor Model & $R_\mathrm{IB}/\Rsun$ & $R_\mathrm{out}/\Rsun$ & resolution ($r\times\theta\times\phi$) & $m_\mathrm{IB}/\Msun$ & $\rphot/\Rsun$ & $\DR/\Rsun$ \\ \hline
RSG1L4.5* & 400 & 22400 &  $384\times128\times256$ & 12.8 & $796$ & $80$ \\ \hline
RSG2L4.9 & 300 & 6700 & $256\times128\times256$ & $10.79$ & $902$ & $200$\\ \hline

\end{tabular}
\end{center}
\caption{Properties of the 3D progenitor models from \citet{Goldberg2021}, including inner boundary ($R_\mathrm{IB}$), outer boundary ($R_\mathrm{out}$), resolution, mass interior to the simulation domain $(m_\mathrm{IB})$,} {photospheric} radius $\rphot$, and span of the fluctuations in the expected radius of SBO $\DR$ (see discussion in \S\ref{sec:FIDUCIAL}). The simulation domain extends from $\theta=\pi/4-3\pi/4$ and $\phi=0-\pi$, with $\delta r/r\approx 0.01$. The naming scheme indicates $\log(L/\Lsun)$. The * denotes the model used in our fiducial explosion.
\label{tab:models}
\end{table*}

To simulate the ejection of the 3D RSG envelope, we drive a strong, initially spherical shock through the \Athena\ models, introduced at $\Rib$. The required time-dependent inner boundary condition is derived from a 1D hydrodynamic simulation of the shock and ejecta evolution for an appropriately scaled thermal-bomb explosion of a 1D RSG envelope.
The 1D explosion is carried out in \MESA\ r-15140\citep{Paxton2011,Paxton2013,Paxton2015,Paxton2018,Paxton2019}, with a modified version of the \texttt{example\_ccsn\_IIp} test suite case, adapted to excise the entire He core and deposit energy only in the H-rich envelope. It is run to 10 days past SBO, yielding $T$, $\rho$, $m$, radiative flux $F_\mathrm{rad}$, and velocity $v_r$ at $\Rib$.

When the 1D shock front has just passed $\Rib$ (at radius $\rshock\approx\Rib+30\Rsun$), we map the 
post-shock $\rho$, $T$, and $v$ to our \Athena\ simulation domain from $\Rib$ to $\rshock$, leaving the 
pre-shock 3D envelope above that location unchanged.
We then demand that the time-dependent boundary condition for $T$, $\rho$, $v$, $m$, and $F_\mathrm{rad}$ at
$r=\Rib$ in \Athena\  match that of the exploded \MESA\ model at the $\Rib$ coordinate at each time step 
thereafter. \deleted{We verified this approach by directly running both \MESA\ and \Athena\ for the same 
spherically symmetric cases, finding agreement.} We discuss this setup and verification in greater 
detail below, in \S\ref{sec:VERIFICATION}. For our fiducial RSG1L4.5 explosion, 
$8\times10^{50}\,\mathrm{erg}$ is deposited into the H-rich envelope, comparable to an $\approx10^{51}$ erg 
explosion of the whole star.


\subsection{Shock {Initialization} and Verification \label{sec:VERIFICATION}}
We now discuss our \MESA-motivated \Athena\ inner boundary explosion scheme, and compare spherical explosions on our 3D grid. The spherically symmetric SN shock problem is well-understood in the strong-shock, radiation-dominated limit. \citet{Matzner1999}
provide an analytic expression for the shock velocity at radius $r$,
\begin{equation}
    v_{\mathrm{sh}}(r)=A\left(\frac{E_\mathrm{exp}}{\Delta m}\right)^{1 / 2}\left[\frac{\Delta m}{\rho_0 r^{3}}\right]^{0.19},
    \label{eq:MMvshock}
\end{equation}
where $E_\mathrm{exp}$ is the explosion energy, $\rho_0(r)$ is the local pre-shock 
density, $\Delta m$ is the mass entrained by the shock, and $A = 0.736$ \citep{Tan2001}. In the radiation-dominated post-shock
plasma ($\gamma = 4/3$) the velocity of the fluid just behind the strong shock front is $v_{\mathrm{fast}}=6v_\mathrm{sh}/7$. The density contrast between the pre- and post-shock material
is $\rho_1/\rho_0=(\gamma+1)/(\gamma - 1)=7$ for $\gamma=4/3$, where a 1 subscript denotes the 
post-shock properties and 0 denotes pre-shock properties \citep{Zeldovich1967}. This expression is valid when radiation is unable to leak out of the shock front ($\tau\gg c/\vsh$).
\citet{Paxton2018} showed excellent agreement between explosions in \MESA\ and these semi-analytic expectations. 

In an exploding star, reverse shocks from core boundaries alter
the final structure and composition of the inner SN ejecta via the Rayleigh-Taylor Instability 
(RTI). This is captured in \MESA\ via the \citet{Duffell2016} RTI prescription. However, we show below that
when studying the propagation of the forward shock through the hydrogen-rich envelope and the properties 
of the outer ejecta shortly after shock breakout, these effects can safely be ignored.
To provide a 1D model which we can
import into \Athena\ as a new bottom boundary condition, we excise the entire He core in the \MESA\ model, and put a thermal bomb in
the innermost $0.2\Msun$ of the H-rich envelope to reach a specified total final 
energy.

As a first verification, we explode the 99em16 progenitor model from \citet{Paxton2018}
in \MESA\ with the full core-envelope structure included, and then again 
with a lower explosion energy and the entire He core excised.
{In both cases, we excise the core with an entropy cut (at 4 kb/baryon for the Fe core and 20 kb/baryon for the He core) before causing the infall to stall (at 400 km for the Fe core and $1\Rsun$ for the He core) and depositing the explosion energy in the innermost $0.2\Msun$ of the ejecta. We then allow the resulting shock to propagate out through the envelope following the \MESA-r15140 test\_suite case \texttt{ccsn\_IIp} (see \citealt{Paxton2018} for discussion).}


\begin{figure}
\centering
\includegraphics[width=\columnwidth]{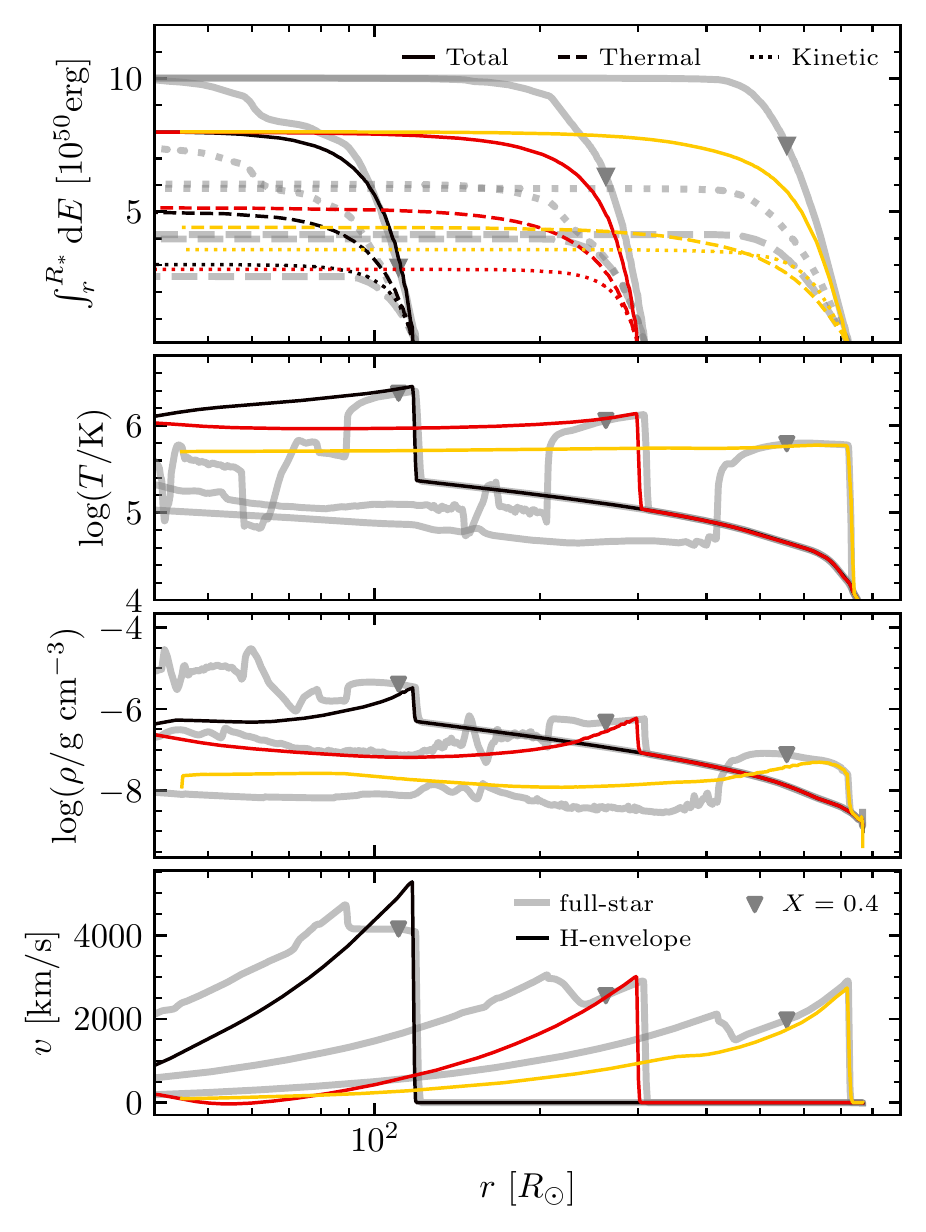} 
\caption{Upper panel: Cumulative energy integrated from the surface for three snapshots from \MESA\ at different times, in a $10^{51}\mathrm{erg}$ full-star explosion (thick grey lines) and a $8\times10^{50}\mathrm{erg}$ explosion of the hydrogen-rich envelope (thin colored lines). Integrated thermal energy (dashed lines), kinetic energy (dotted lines), and total energy (solid lines) are shown. Temperature (second panel), density (third panel), and velocity (fourth panel) are shown for the same three snapshots. Grey triangles approximate the He core boundary where $X=0.4$ in the full-star explosions. 
}
\label{fig:RTIhack}
\end{figure}

\begin{figure*}
\centering
\includegraphics[width=0.48 \textwidth]{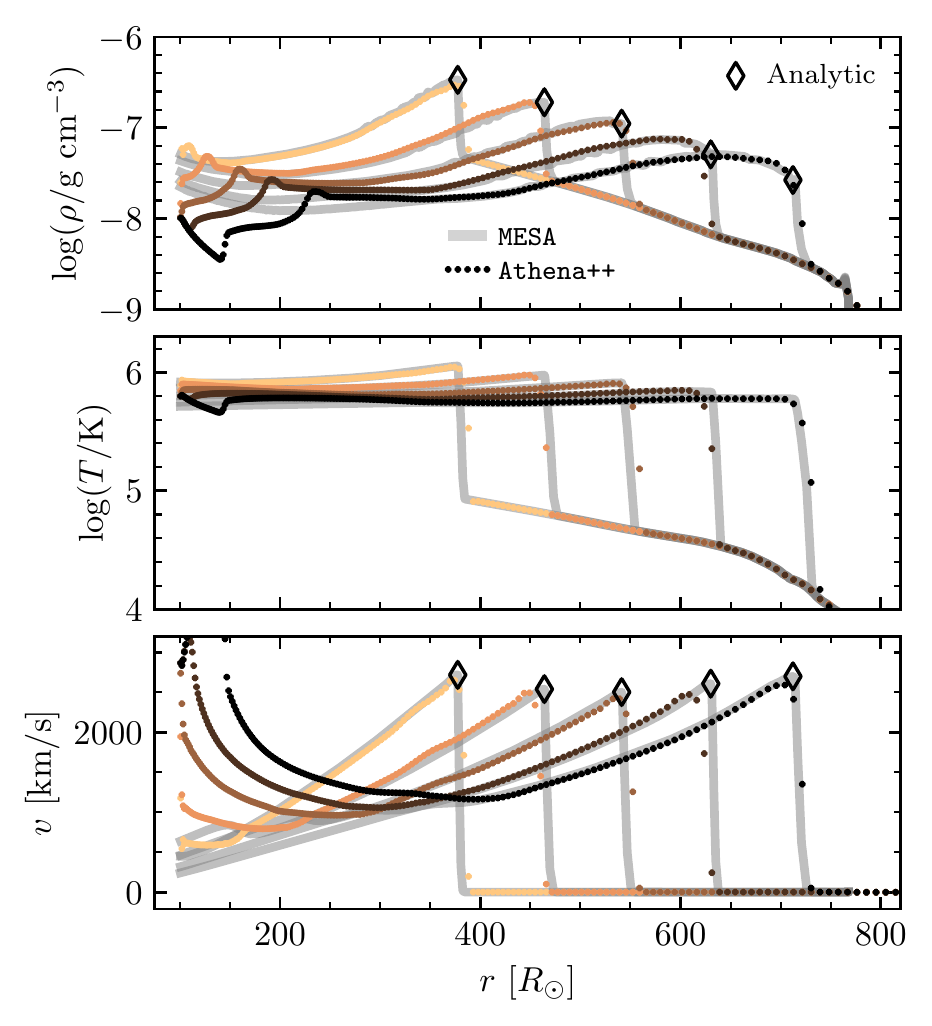} %
\includegraphics[width=0.48 \textwidth]{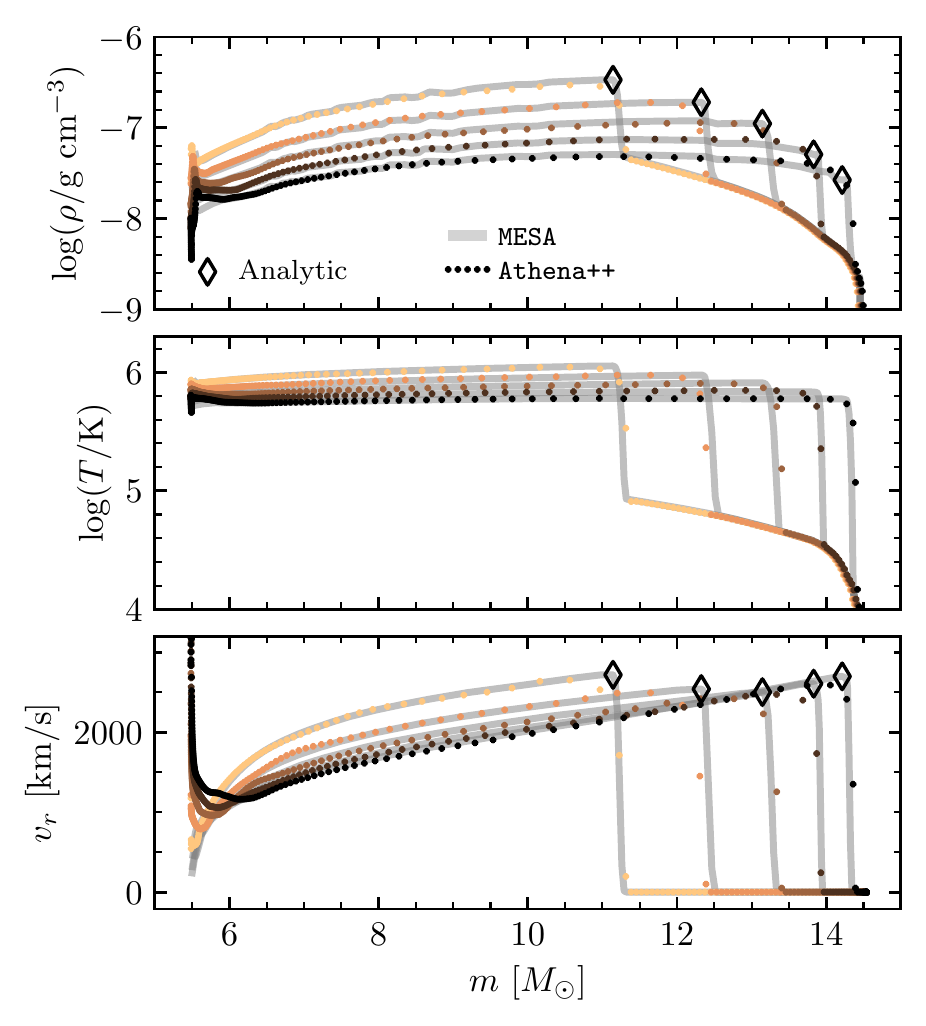} %
\caption{{Density (upper panels), Temperature (middle panels), and velocity (lower panels) 
profiles as a function of radius (left column) and mass coordinate (right column) for the same}
{spherical} shock in \MESA\ (thick grey lines) and \Athena\ (colored points). 
Point spacing indicates the radial grid resolution in \Athena.
Black diamonds show analytic expectations for post-shock density $\rho_1=7\rho_0$ and and 
fluid velocity $v_\mathrm{fast}=6v_\mathrm{sh}(r)/7$ using values from \MESA. }
\label{fig:shockpropagation}
\end{figure*}

\begin{figure}
\centering
\includegraphics[width=\columnwidth]{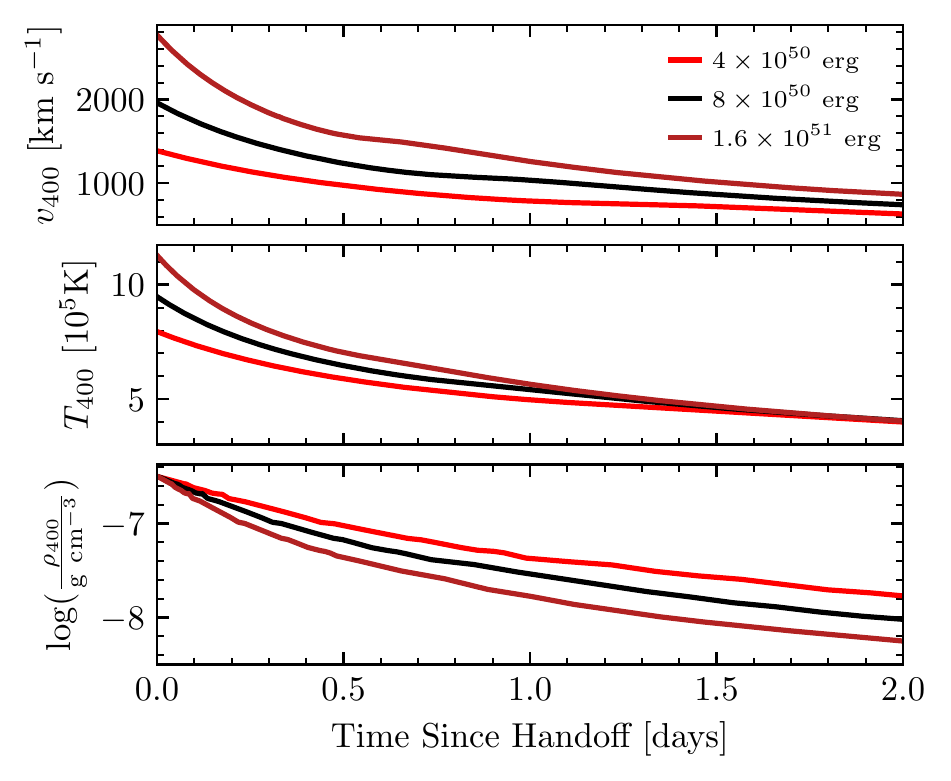} %
\caption{Velocity, temperature, and density at
$r=r_\mathrm{IB}=400R_\odot$ in \MESA\ starting at the time of handoff to \Athena\ for the 3 explosion energies discussed in \S\ref{sec:OBSERVABLES}.}
\label{fig:timeIBC}
\end{figure}

A comparison of the shock properties from the two \MESA\ simulations is shown in Figure \ref{fig:RTIhack}. 
Snapshots are selected such that the shock fronts are at approximately 
the same radius. The grey lines show profiles of the $10^{51}$ erg full-star explosion, while the 
thin colored lines show the equivalent 8.0$\times10^{50}$ erg shock in the H-envelope. 
We see {good} agreement in the post-shock temperatures and velocities, and satisfactory agreement in 
the outer post-shock density, especially at late times as the shock wave has had sufficient 
time to propagate. The post-shock material is nearly isothermal, and the shock velocity 
behaves as predicted by Equation \eqref{eq:MMvshock}. 
In the full-star explosion, roughly 80\% of the energy is deposited within the H-rich envelope, which motivates the choice of an $8.0\times10^{50}\mathrm{erg}$ energy deposition to compare to a $10^{51}\mathrm{erg}$ full-star explosion and for our fiducial 3D explosion.
Although the reverse shock in the full-star explosion accounts for $<20\%$ of the explosion energy, it is 
not relevant to the surface material nor the propagation of the forward-shock.

To ensure that this spherically symmetric explosion is also captured in 
\Athena, we import the H-only explosion from \MESA\ to \Athena\  at a time 
when the shock front is at $r\approx300R_\odot$, 0.435 days after the onset of the explosion 
in \MESA. We use a similar spherical polar grid as in our 3D RSG model, with periodic boundary conditions in
$\theta$ and $\phi$ ($\theta=\pi/4-3\pi/4$, $\phi=0-\pi$), a reflective-velocity inner boundary (at 
$100 R_\odot$), and an outflow outer boundary beyond $r=2200R_\odot$. 
Figure \ref{fig:shockpropagation} compares the resulting shock profiles in \MESA\ (thick grey lines) 
and \Athena\ (colored dotted lines). Snapshots are again selected 
such that the shock fronts are at approximately the same radius at times 
$t=$0.19, 0.41, 0.67, 0.85, and 1.07 days since handoff in \MESA\ and 
$t=$0.19, 0.39, 0.64, 0.83, and 1.08 days since handoff in \Athena.
The radial mesh resolution in \Athena\ is given 
by the dotted points.
 {The \Athena\ data shown are angle-averages, though individual radial rays show the same results for spherical explosions.}
The two software instruments agree well with each other and 
with the analytic expectations, with differences
arising only due to the \Athena\ resolution near the shock front and reflection of some
low-density material off of the inner boundary. These differences do not impact the shock 
propagation or the early post-shock-breakout evolution of the outer envelope. This agreement 
gives us confidence in our use of \Athena\ for the next stage of our exploration.

The inner boundaries of our 3D RSG envelope simulations are
at $r=300-400R_\odot$ rather than $r=100R_\odot$ as above, which means the scheme discussed above 
must be modified to explode the 3D envelopes. As seen in Figure \ref{fig:RTIhack}, 
a significant portion of the mass and explosion energy passes through the $r=400R_\odot$ as the explosion 
approaches shock breakout. We therefore developed a second explosion scheme on the exact same grid as our 
fiducial 3D envelope model, but instead of a fixed inner boundary with specified initial 
conditions, we specify the time-dependent fluid velocity $v$, density $\rho$, temperature
$T$, and co-moving radiative flux $F_\mathrm{rad}$ in the ghost zones at the $r=400R_\odot$ inner 
boundary to match a \MESA\ model at the $r=400R_\odot$ coordinate at each timestep. This allows energy and mass to be fed into the model through the inner boundary in order to power the shock. 

For this, we must choose a \MESA\ model which matches the average properties of the
RSG1L4.5 envelope model at the $r=400R_\odot$ coordinate. We thus select a \MESA\ model selected from the \cite{Goldberg2020} grid of progenitors with pre-shock $\rho(r=400\Rsun)$, $T(r=400\Rsun)$, and $m(r=400\Msun)$ approximately matching the shell-averaged values at the $r=400R_\odot$ in the fiducial RSG1L4.5 envelope. The chosen model has a 
progenitor mass of $19M_\odot$, mixing length in the H-rich envelope $\alpha_H=3.0$, modest 
wind $\eta_\mathrm{wind}=0.2$, no rotation ($\omega/\omega_\mathrm{crit}=0$), no overshooting 
($f=f_0=0$), {and metallicity $Z=0.02$.} The mass and radius of this \MESA\ RSG progenitor model at the time of core-collapse are $18.5M_\odot$ and $659R_\odot$.

As a first test of this scheme, we populate the \Athena\ grid with data from the profile of the H-envelope only explosion in \MESA\ from $r=400R_\odot$ outward at the time when the shock radius in \MESA\ is at $r_\mathrm{shock}=426\Rsun$, near but outside the inner boundary in \Athena. At the time of handoff to \Athena, the enclosed mass {below} $r=400R_\odot$ is 12.7$M_\odot$. 
Figure \ref{fig:timeIBC} then shows our time-dependent inner boundary condition on $v$, $T$, and $\rho$ taken from the 
$r=400R_\odot$ coordinate in \MESA, with the time of handoff identified when $\rshock=426\Rsun$. The radiative luminosity, which is also passed to \Athena, is small compared to the advective luminosity and the sign
is negative, as the temperature gradient is positive in the post-shock material. The comoving radiative luminosity can also be estimated from $F_\mathrm{r}=\frac{1}{3}(c/\kappa\rho)aT^4/400R_\odot$, where $\rho$, $T$, and $\kappa$ are taken at the 400$R_\odot$ coordinate, which gives $\log(F_{r,400}/\mathrm{erg\,s^{-1}\,cm^{-2}})\approx$13.3, 13.0, and 13.6 for the fiducial (black), low-energy (light red) and high-energy (dark red) explosions.

\begin{figure}
\centering
\includegraphics[width=\columnwidth]{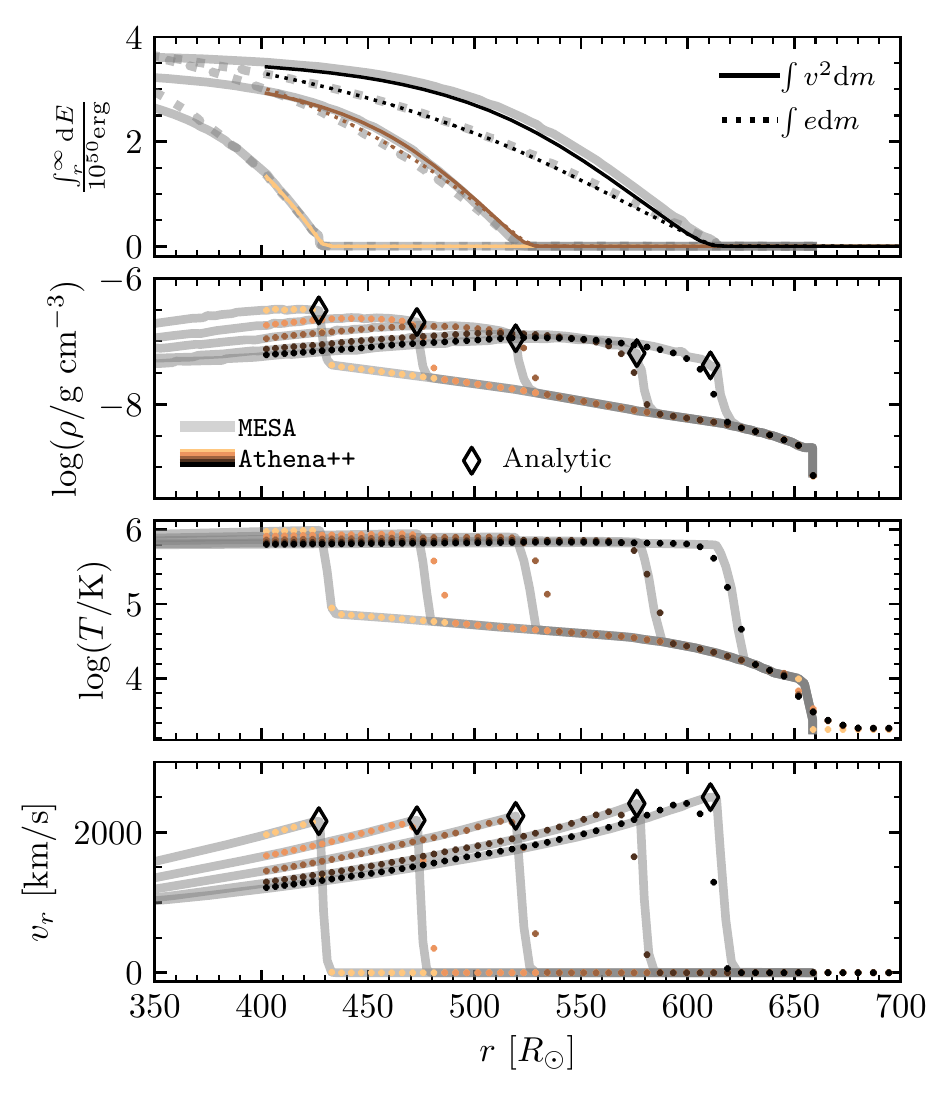} %
\caption{Upper panel: Cumulative kinetic (solid lines) and thermal (dashed lines) energy integrated from the surface to radius $r$ for different shock locations in \MESA\ (thick grey lines) and \Athena\ (colored lines/points) using the time-dependent boundary condition at $r=400\Rsun$.
Density (second panel), temperature (third panel), and velocity (fourth panel) 
profiles are also shown. Point spacing in the lower three panels indicates the radial grid 
resolution in \Athena. Black diamonds show analytic expectations for post-shock density 
$\rho_1=7\rho_0$ and and fluid velocity $v_\mathrm{fast}=6v_\mathrm{sh}(r)/7$ using values from \MESA. }
\label{fig:shockprop2}
\end{figure}

Figure \ref{fig:shockprop2} compares the shock propagation and energetics in \MESA\ and 
\Athena\ for the $8\times10^{50}$ erg energy {deposition}, at times $t=$0.0, 0.15, 0.29, 0.46, and 0.56 days after 
handoff (which occurs 0.96 days after explosion) in \MESA\ and $t=$0.0, 0.15, 0.30, 0.44, and 0.54 days since handoff in \Athena. 
The upper panel only shows every other profile for clarity, and the \Athena\ grid is given by the rounded points in the lower 3 panels. 
We see excellent agreement between shock properties in \MESA\ and \Athena, with 
discrepancies arising primarily due to the slight time differences between the \MESA\ and \Athena\ profile output. At the time of
shock breakout, nearly 80\% of the total shock energy is contained within the \Athena\ simulation 
domain, nearly equipartitioned between kinetic and thermal energy, with thermal energy accumulated 
deeper in the ejecta. This continued agreement further bolsters our confidence in the use of \Athena\ 
to explore the 3D problem.

\subsection{The 3D shock}
To {drive} this explosion in the 3D envelope, we start with the 3D RHD \Athena\ envelopes at the end of the simulation run described in \citet{Goldberg2021}.
We then populate the innermost portion of the \Athena\ simulation domain with the
post-shock $\rho$, $T$, and $v$ values from the exploded \MESA\ model between $\Rib$ to $\rshock$ when $\rshock\approx\Rib+30\Rsun$, leaving the 
pre-shock 3D envelope above that location unchanged. We then demand that the time-dependent $T$, $\rho$, $v$, $m$, and $F_\mathrm{rad}$ at $r=\Rib$ in \Athena\  match that of the exploded \MESA\ model thereafter. 
The mass in our simulation domain in the fiducial explosion is 4.5$\Msun$ at the time of the peak bolometric luminosity ($\Lbol$) in the SBO, which extrapolates to $\Mej=12.7\Msun$ accounting for the limited solid angle, which is a combination of the initial mass within the simulation domain and the mass fed in by this time-dependent boundary condition.

\begin{figure}
\centering
\includegraphics[width=\columnwidth]{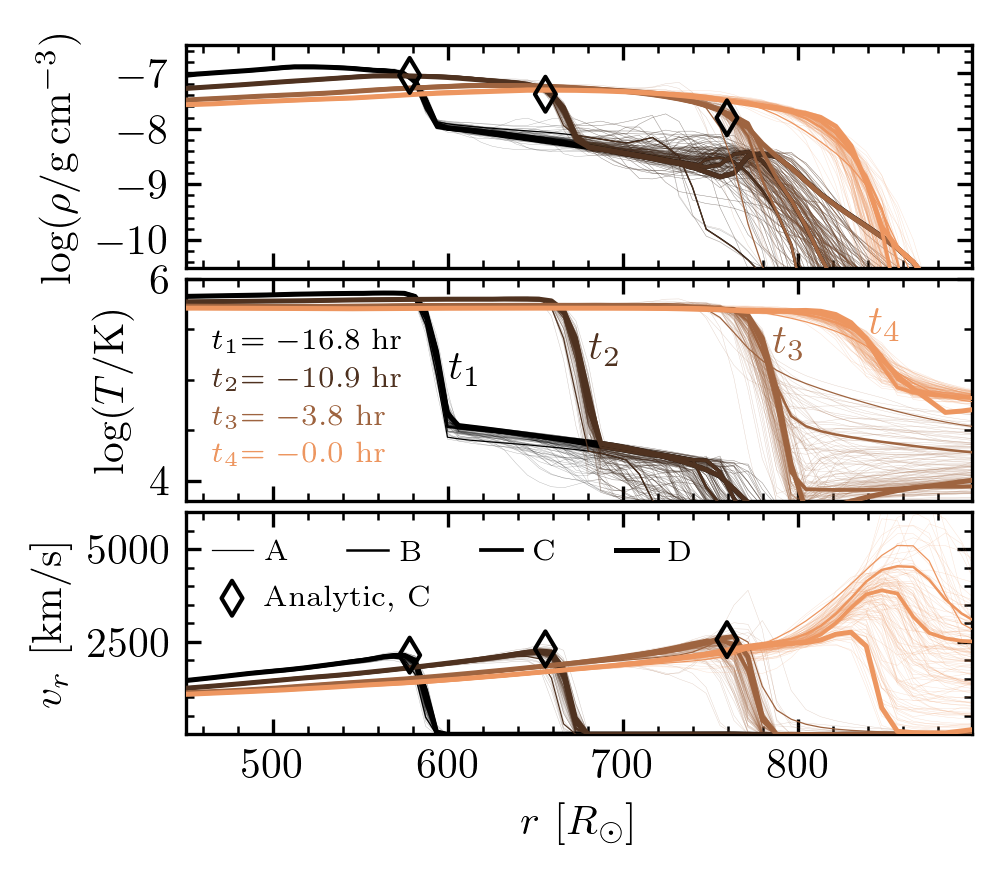} %
\caption{{Density (top),} temperature (middle), and $v_r$ (bottom) radial profiles for our fiducial 3D explosion along 128 rays (thin, faint lines) uniform in ($\theta,\phi$), with four lines of sight with different expected SBO times emphasized (thicker solid lines), at different times. Where applicable, the analytic expressions {are shown for the density, $7\rho_0$, and} fluid velocity near the shock front, $v_\mathrm{fast}=6\vsh/7$ where $\vsh$ is given by Eq.~\ref{eq:MMvshock} using values for point C.}
\label{fig:3DshockProfiles}

\end{figure}

The shock propagation is shown in Fig.~\ref{fig:3DshockProfiles} for our fiducial explosion with $\Mej\approx12.7\Msun$ and $\Eexp=0.8\times10^{51}\mathrm{erg}$. We show 128 radial rays equally distributed across the stellar surface, and highlight four rays corresponding to {different topographical features of the 3D stellar surface: a ``valley" (A), a ``hillside" (B), a ``plateau" (C), and a ``mountain" (D) on the stellar surface, with shock breakout expected to occur approximately in alphabetical order. We discuss this expectation in greater detail in \S\ref{sec:FIDUCIAL}}. Times are labelled as time to the maximum observed $\Lbol$ {integrated over the simulation angular domain.} The \citet{Matzner1999} prediction for the fluid velocity is shown where valid, using values of $\rshock$ and $\rho_0$ along point C, displaying good agreement. We verified agreement along all radial rays, and the variety in $v_r$ at the shock front at any given time is consistent with the variation in $\rho_0$ and $\rshock$. Additionally, we confirm that the post-shock $\rho_1=7\rho_0$ as expected analytically.

\begin{figure}
\centering
\includegraphics[width=\columnwidth]{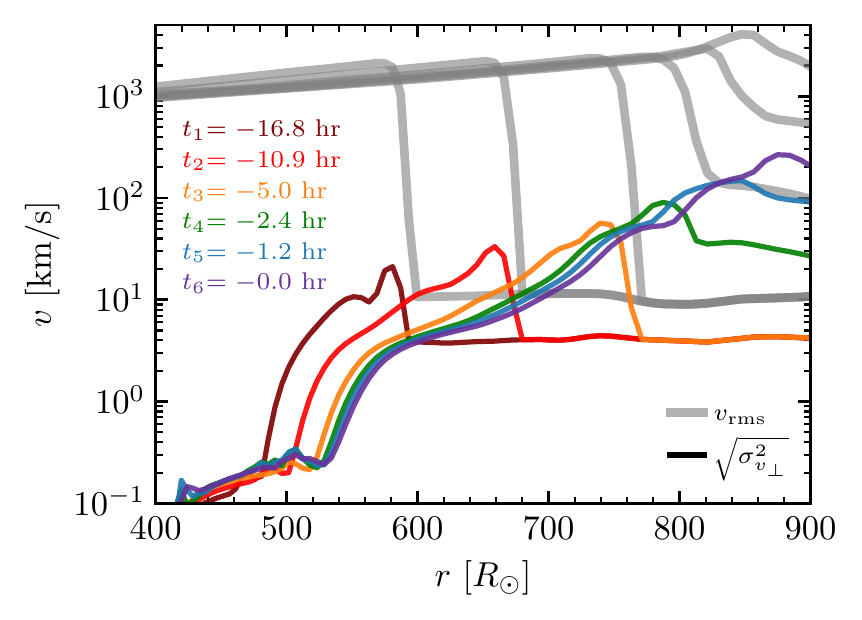} %
\caption{Tangential velocity dispersion (rainbow lines) compared to the mass-weighted rms velocity (thick grey lines) for 6 snapshots of our fiducial explosion.}
\label{fig:Dispersion}
\end{figure}

\begin{figure*}
\centering
\includegraphics[width=1.6\columnwidth]{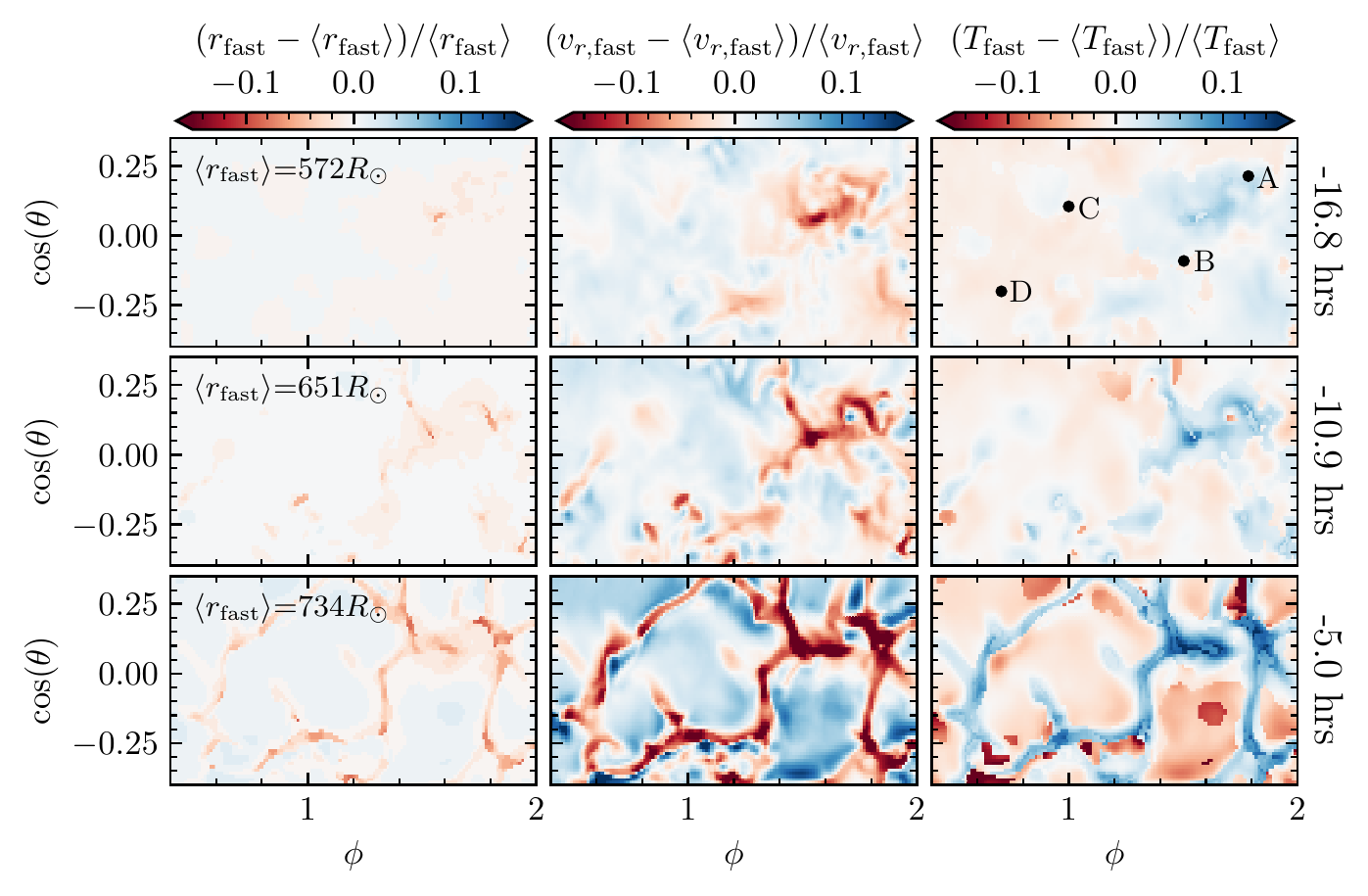} %
\caption{Corrugation of the shock front as it travels through the convective H-rich envelope prior to SBO. Color indicates fluctuations in radial coordinate (left column), $v_r$ (middle), and $T$ (right) of the fastest-moving material at each angular location. The approximate radius of the shock front is labeled in the left column and the time before peak $\Lbol$ is labeled on the right.}
\label{fig:Fluctuations}
\end{figure*}

Prior to the explosion, the convective velocity {($\vc$)} fluctuations are a few to $20$km/s. As $\vsh>2000\mathrm{km\,s^{-1}}\gg\vc$, the resulting change in the shock frame is only at the 1\% level. As the shock passes, the post-shock energy is split in near-equipartition between the radiation energy density and kinetic energy density of the fluid, and the kinetic energy density of the fluid is dominated by {the radial velocity} $v_r$. 
Although the radial motion of the shock front dominates the kinetics, the 
impact of the shock on convective-like fluctuations can be quantified by the tangential velocity dispersion, $\sigma^2_{v_\perp}=\sum_{i}(v_{\perp,i}\intdm_i-\langle v_{\perp,i}\intdm_i\rangle)^2/\sum_i(\intdm_i)$, where $v_\perp^2=v_\theta^2+v_\psi^2$, $\intdm_i$ is the mass in zone $i$ and the sum is over all zones in a given radial shell. This is shown in Fig.~\ref{fig:Dispersion}, which compares the characteristic tangential velocity dispersion (colored lines) to the forward-shock velocity (grey lines), $\vrms=\sqrt{v_r^2+v_\theta^2+v_\phi^2+}\approx|v_r|$, where $\vrms$ is calculated as the kinetic energy in each radial shell divided by the shell mass. The underlying stellar convective dispersion around $4$km/s can be seen in the pre-shock material. Due to the envelope inhomogeneities, the velocity dispersion grows as transverse pressure gradients accelerate the fluid; however there is insufficient variety in the shock arrival time for these fluctuations to grow appreciably, and the tangential velocity dispersion remains an order of magnitude below the shock velocity.

As the explosion progresses, the shock front begins to corrugate from differing density profiles seen along each radial ray. Fig.~\ref{fig:Fluctuations} shows these growing inhomogeneities in shock temperature, velocity, and radius in progressive snapshots before SBO for a zoom-in patch of the stellar surface, {compared to the angle-averages (denoted $\langle\cdots\rangle$).}
Quantities are shown at the radial coordinate of the fastest-moving material at each angular location (denoted $r_\mathrm{fast}$), and the radial velocity and radiation temperature $T_r=(E_r/a)^{1/4}$ at that coordinate are denoted $v_{r,\mathrm{fast}}$ and $T_\mathrm{fast}$ respectively. 
Locations A, B, C, and D, are labeled. {As the shock propagates further through the envelope, the shock radius begins to vary, but before breakout remains at the level of} $\approx\pm3\%$. Temperature fluctuations {reach} the 10\% level, and velocity fluctuations {grow to} the 15\% level.

SBOs in aspherical axisymmetric explosions have been considered in prior works, primarily in the case of oblique shock breakout in more compact (e.g. blue supergiant) stellar sources \citep[e.g.][]{Suzuki2016,Afsariardchi2018}, as well as semi-analytically \citep{Linial2019,Irwin2021}. Some 3D simulations of the core-collapse explosion itself have examined shock propagation all the way up to SBO with neutrino-powered core-collapse explosions (e.g. \citealt{Wongwathanarat2015,Stockinger2020,Sandoval2021,Kozyreva2022}) in RSG envelopes coming from 1D stellar models; in those works, asymmetric shocks, which sphericalize as they propagate, are introduced by the explosion mechanism. In contrast, all effects discussed here are introduced by the 3D convective envelope itself.

\subsection{Measuring the Bolometric Luminosity}

We calculate $\Lbol$ as the integrated $F_r$ passing through our simulation domain at fixed radius scaled to the full $\Omega=4\pi$. We choose 2700$\Rsun$ for a representative location, and we confirmed that the lightcurve properties are independent of this choice. 
Because an observer sees light coming from the whole star at once, not just light that travels along radial rays, 
this representative location, which has a horizon encompassing $95\%$ of the solid angle of the $r\approx820\Rsun$ SBO surface, is convenient for estimating the variation in $\Lbol$, by taking the bolometric flux at different angular locations. When showing fluid properties, 
time 0 is when the star is emitting radiation that corresponds to the peak in $\Lbol$. 
This is another motivation to use $2700\Rsun$ rather than, e.g., $r=2500\Rsun$ or $3000\Rsun$, as the light travel time from the star to the representative location is commensurate with the time sampling of the 3D simulation output.


\section{The 3D breakout}

\begin{figure}
\begin{center}
\includegraphics[width=\columnwidth]{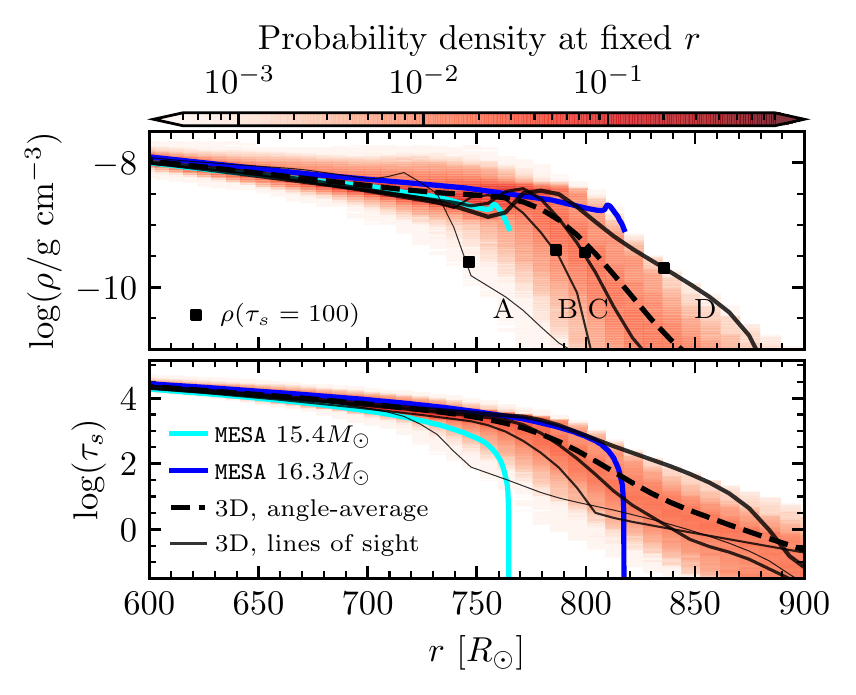} 
\caption{Profiles of density (upper) and $\taus$ (lower) for our pre-explosion snapshot of the fiducial 3D RSG1L4.5 model, compared to 1D profiles \MESA\ models with ZAMS masses of $16\Msun$ (cyan) and $17\Msun$ (blue) {and final stellar masses of $15.4\Msun$ and $16.3\Msun$ respectively}. Red colors indicate volume-weighted probability of finding a fluid element at a given radial coordinate with a given ($\rho,~\taus$). Dashed black lines give the angle-averages $\langle\rho\rangle,~\langle\taus\rangle$, and solid black lines show radial profiles along locations A, B, C, and D, with thicker lines indicating larger $r(\taus=100)$. Squares in the upper left panel indicate $\rho$ where $\taus=100$.}
\label{fig:Rho3Da}
\end{center}
\end{figure}

Radiation escapes ahead of the shock when the shock reaches an optical depth $\tausbo=\,c/\vsh$ \citep{Lasher1979}.
The escaping radiation ionizes the pre-shock material and the opacity thereafter can be well-approximated by electron-scattering, $\kes=0.32\mathrm{\,cm^2/g}$. The radial scattering optical depth ($\taus$) is then related to the column depth $y=\int_r^\infty{\rho\intd r'}$ as $\taus\equiv\kes\y$, and the breakout occurs where $\taus\ltapprox\tausbo$.
In the outer layers, 3D simulations reveal scale heights
which are significantly larger than traditional 1D hydrostatic models, likely owing to turbulent pressure \citep{Chiavassa2011a}. A low-density `halo' of material out to a few hundred $\Rsun$ past the photosphere is also present above a bulbous surface with large-scale plumes and order-of-magnitude density fluctuations spanning tens to hundreds of $\Rsun$. 
Fig.\ref{fig:Rho3Da} shows density ($\rho$) and $\taus$ profiles immediately prior to explosion for RSG1L4.5. At $\rphot=796\Rsun$, the shell-averaged density is $\langle\rho\rangle=6.9\times10^{-10}\mathrm{\,g\,cm^{-3}}$ and column depth is $\langle\y\rangle=780\,\mathrm{\,g\,cm^{-2}}$, with 80\% of the material between $\rho=4.0\times10^{-11}-1.9\times10^{-9}\mathrm{\,g\,cm^{-3}}$ and $y=50-2100\mathrm{\,g\,cm^{-2}}$. Before the explosion, the opacity in this outer material is $\kappa\sim10^{-3}\mathrm{\,cm^2\,g^{-1}}$.
For RSG2L4.9, not shown, $\rphot=902\Rsun$, and $\langle\rho\rangle=7.1\times10^{-10}\,\mathrm{g\,cm^{-3}}$,
with 80\% between $\rho=3.4\times10^{-11}-1.8\times10^{-9}\,\mathrm{g\,cm^{-3}}$.

\label{sec:FIDUCIAL}

\begin{figure}
\includegraphics[width=\columnwidth]{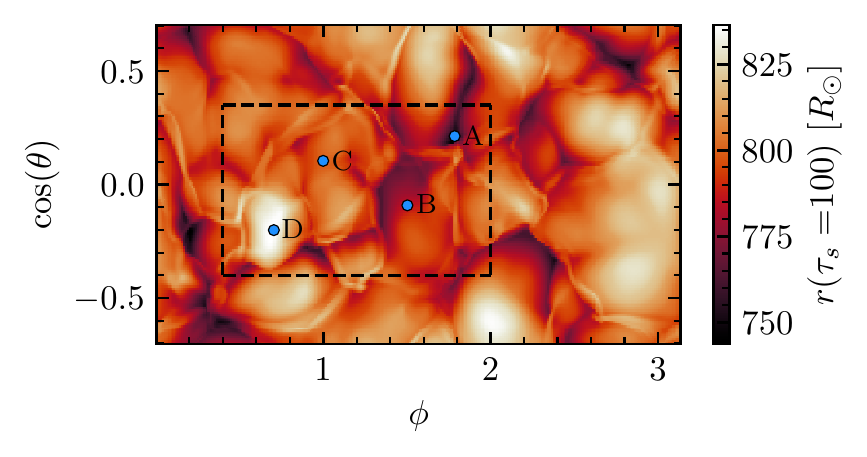} 
\caption{Topographical map of the radius where $\taus=100$ along each radial direction. Four characteristic lines of sight (A,B,C,D) are labeled, and the dashed box indicates the zoom-in region for Figs.~\ref{fig:Fluctuations} and \ref{fig:sbopanels}.}
\label{fig:Rho3Db}
\end{figure}

Our fiducial explosion of RSG1L4.5 generates a shock with $\vsh\approx3000$km/s approaching breakout, corresponding to $\tausbo\approx100$. 
The $\taus=100\approx\tausbo$ surface, spanning $\DR\approx80\Rsun$, is shown in Fig.\ref{fig:Rho3Db}, 
{which would correspond to a horizontal slice through the bottom panel of Fig.~\ref{fig:Rho3Da}}. 
The dashed box shows the characteristic patch which we zoom into, with four points at different topography labeled: a ``valley" (A), a ``hillside" (B), a ``plateau" (C), and a ``mountain" (D). 
{The radial profiles corresponding to each of these locations are shown as black lines in Fig.~\ref{fig:Rho3Da}.}
The location of SBO often lies outside the traditionally-defined $\rphot$, at characteristically lower densities. 
In contrast, 1D RSG models with a barren photosphere (i.e. no circumstellar material; blue lines in Fig.\ref{fig:Rho3Da}), show $\taus$ 
plummeting from 1000 to 1 over $\approx$7$\Rsun$ ($\approx$1\% of the stellar radius) and $\rho\approx10^{-9}\,\mathrm{g\ cm^{-3}}$ for 
the lowest-density material, about $5\times$ higher than we find at $\tausbo$ in 3D models.

\begin{figure}
\begin{center}
\includegraphics[width=\columnwidth]{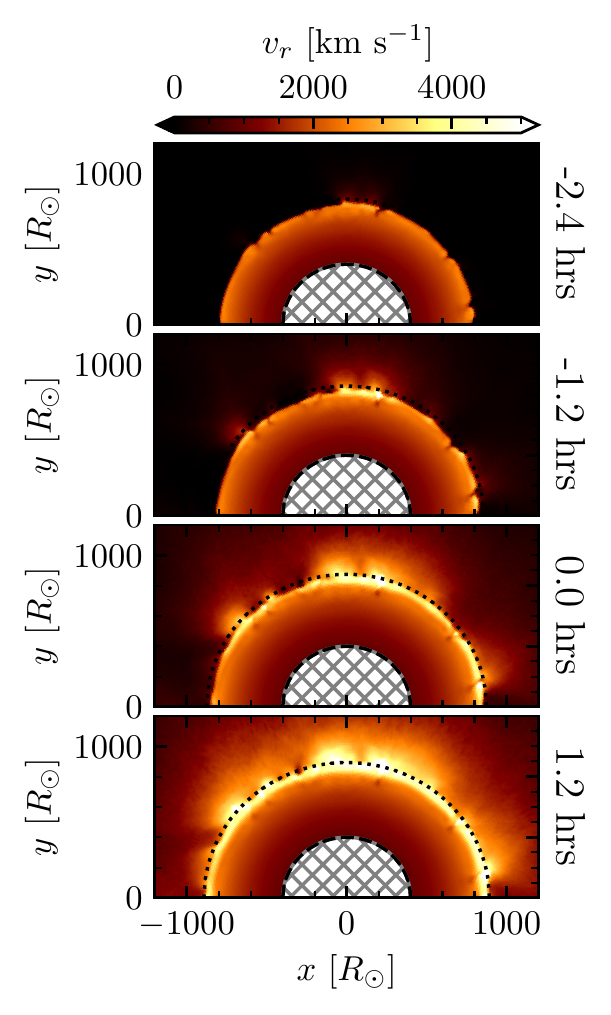}
\caption{Time lapse of radial velocities (orange colors) for equatorial slices of our fiducial explosion, with time snapshots indicated relative the lightcurve peak. The dotted line denotes $\rphot$, which moves outwards as the surface expands, and the simulation inner boundary is shown by grey thatches.}
\label{fig:breakoutvr}
\end{center}
\end{figure}

\begin{figure*}
\begin{center}
\includegraphics[width=\textwidth]{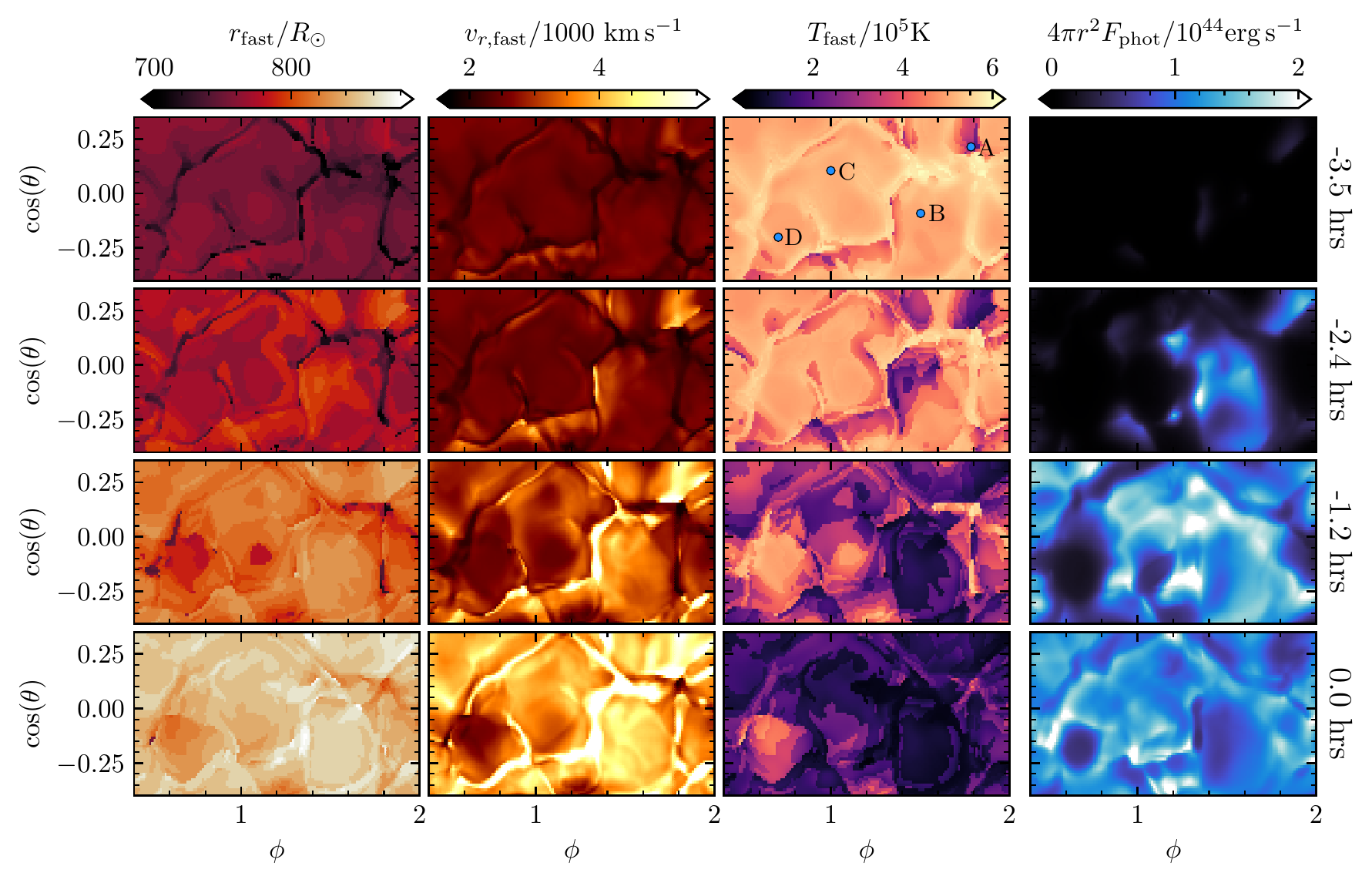}
\caption{
Left to right: Radial coordinate (first column), radial velocity (second column), and temperature (third column) taken at the fastest-moving ejecta along each radial line of sight, as well as $4\pi r^2F_{r}$ at the photospheric surface (fourth column). The solid-angle spans the dash-enclosed region of Fig.\ref{fig:Rho3Db}, for our fiducial explosion at different times approaching the peak in the bolometric luminosity (labeled to the right of each row). Reference locations A, B, C, and D are shown in the ($T_\mathrm{fast}$, -3.5 hours) panel.}
\label{fig:sbopanels}
\end{center}
\end{figure*}

Two timescales are relevant to the observed SBO duration from a 3D turbulent star. The first is the local radiation diffusion time along a radial ray at the moment of SBO \citep{Katz2012}
\begin{equation}
    \tdiff\approx\frac{\hrho}{c}\tau\approx\frac{c}{\kappa \rho_0\vsh^2},
    \label{eq:tdiff}
\end{equation}
where $H_\rho$ is the local (radial) density scale height and the second expression eliminates $\hrho$ by equating the first expression and the time for the shock to cross $\hrho$ ($\hrho/\vsh$; i.e. the breakout condition) and expressing $\tau$ as $\kappa\rho\hrho$ near the surface.
For our fiducial model where $\taus=100$, $\rho_0$ ranges from $1.25-3\times10^{-10}\mathrm{g\,cm^{-3}}$, yielding $\tdiff\approx1-2.2$ hours, $3-10\times$ longer than $\tdiff$ in 1D models with barren photospheres \citep{Shussman2016}. 
The second timescale, intrinsic to 3D stars, is the time it takes the shock to reach all of the fluid elements spanned by the $\DR\approx80\Rsun$ corrugation at the surface, 
\begin{equation}
    \tcross\approx\DR/\vsh.
    \label{eq:tcross}
\end{equation}
For $\DR=80\Rsun$ and $\vsh=3000-5000\,\mathrm{km\,s^{-1}}$ {as is typical in the broken-out rays for the fiducial explosion seen in Fig.~\ref{fig:3DshockProfiles}}, this is 5 hours to 3 hours. For the 3D stellar progenitor, this timescale dominates $\tdiff$, and most importantly, the light-travel time across the star, $R/c$, which is $\approx0.5$ hours for $R\approx800\Rsun$.

This timing spread can be seen as the shock reaches the inhomogeneous outer layers and accelerates down the steeper outer density gradient. 
Valleys on the stellar surface (like point A) are shocked first, and mountains (like point D) are shocked later.
Fig.\ref{fig:breakoutvr} shows the radial velocity, $v_r$, for equatorial slices of our fiducial explosion at four snapshots, with $t=0$ at the bolometric luminosity ($\Lbol$) peak. Most of the motion of the shock remains radial.
When the shock velocity is $v_r\approx3000\,$km/s (see Fig.~\ref{fig:Dispersion}), the transverse velocities are $\vperp\approx300$km/s; even in broken-out layers where fluctuations begin to span an order of magnitude in $\rho$ and the shock accelerates to $4000-10000$km/s, $v_{\perp}$ stays below 1000 km/s.  

Fig.\ref{fig:sbopanels} shows the radial coordinate of the fastest-moving material at 
each angular location (denoted $r_\mathrm{fast}$), the radial velocity and radiation temperature $T_r=(E_r/a)^{1/4}$ at that coordinate (denoted $v_{r,\mathrm{fast}},T_\mathrm{fast}$ respectively), and the local
flux at the first location where $F_r/E_r=1/3$ inwards along each radial ray. We focus on a zoom-in (dashed box in Fig.\ref{fig:Rho3Db}) of our 
fiducial $0.8\times10^{51}\,\mathrm{erg}$ explosion of the RSG1L4.5 model. Times are relative to peak $\Lbol$. 
We do not find sufficient obliquity that material accelerated in transverse 
directions is able to wrap around and reach the outer layers (see discussions in \citealt{Irwin2021}) before the forward shock arrives. Rather, the 
``valleys" (A:$\phi=1.78,\cos(\theta)=0.21$)
have already undergone breakout and cooled over $\tdiff$, before the ``mountains" are hit (D:$\phi=0.7,\cos(\theta)=-0.2$). 
This manifests in lower $T_\mathrm{fast}$, higher $v_r$, and higher $F_\mathrm{phot}$ for point A at -3.5 hours, whereas for point D, the shock front retains its heat and $v_{r,\mathrm{fast}}$ remains lower, not yet undergoing breakout even at the time of peak $\Lbol$. 
Another intriguing outcome is that as the material in the valleys is shocked first and accelerated at an earlier time it gets to larger radii first, inverting the topography of the surface of maximum velocity, evident in the lower left panel of Fig.~\ref{fig:sbopanels}.

\section{Observed properties of the 3D SBO \label{sec:OBSERVABLES}}

\begin{figure}
\begin{center}
\includegraphics[width=\columnwidth]{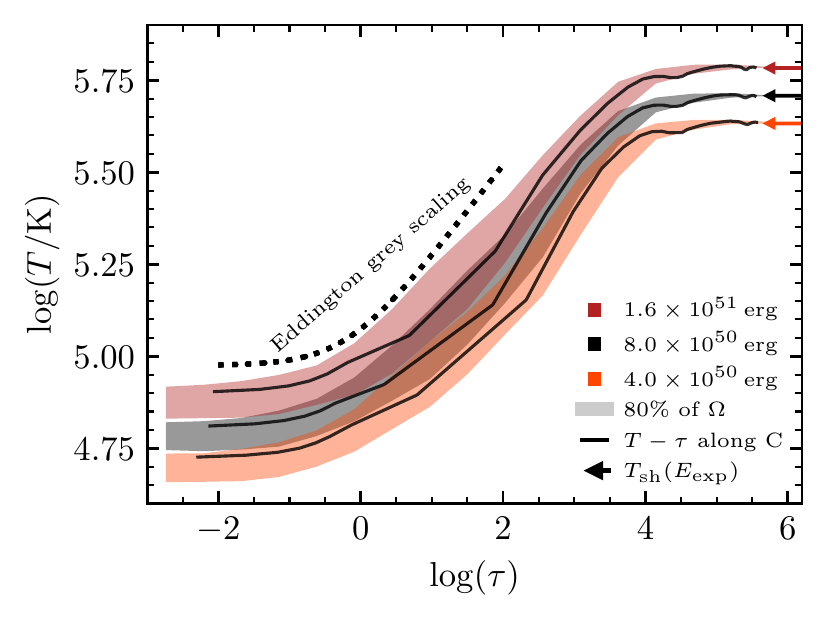}
\caption{$T-\tau$ relations for 3 explosions varying $\Eexp$,
taken at peak $\Lbol$. Shaded regions show relations for 80\% 
of the solid-angle, and arrows indicate the asymptotic shock 
temperature predicted by Eq.~\ref{eq:TofE}. Thin black lines show $T-\tau$ relations at location C. 
The scaling for an Eddington grey atmosphere assuming constant flux is shown for reference.}
\label{fig:ttau}
\end{center}
\end{figure}

\subsection{Temperature Structure}

In the 1D picture, the observed temperature is set by the energy density at the breakout location $T\approx(\rho\vsh^2/a)^{1/4}$
up to a factor of order unity accounting for thermalization of the radiation and gas \citep{Nakar2010,Sapir2013}. For $\vsh=3000-5000$km/s and $\rho=1.25-3\times10^{-10}\mathrm{g\,cm^{-3}}$, this would predict $\log(T/\mathrm{K})=5.3-5.5$. As the post-shock temperature profile is nearly constant 
throughout the deep envelope, a similar estimate is $4\pi\rshock^3a\Tsbo^4/3=\Eexp/2$, with $\rshock$ taken at the average radius where $\taus=c/\vsh$, 
or
\begin{equation}
    \Tsbo=\left(\frac{3\Eexp}{8\pi \rshock^3a}\right)^{1/4}. 
    \label{eq:TofE}
\end{equation}
Fig.~\ref{fig:ttau} shows radiation temperature as a function of
(radial) optical depth $\tau=\int_r^\infty\kappa\rho\intdr$, for 
snapshots corresponding to the peak in $\Lbol$ for 3 explosions of RSG1L4.5 at 
different $\Eexp$. Shaded regions indicate the spread in $T-\tau$ relations for 80\% 
of $\Omega$, and the arrows show the prediction for the asymptotic shock 
temperature from Eq.~\ref{eq:TofE} using the average shock front location across the 
breakout surface, $\rshock=820\Rsun$. 
The thin black lines show $T-\tau$ relations along an intermediate ray (point C) which has recently been hit by the forward shock. 
The radiation and gas temperatures are approximately thermalized in 
the material which has already broken out, and the profile follows the Eddington grey atmosphere assuming constant flux, 
$T^{4}=\frac{3}{4} T_{\mathrm{eff}}^{4}\left(\tau+\frac{2}{3}\right)$ (\citealt{Rybicki1986}, dotted line). When $\tdiff\ll\tcross$, we expect a diversity of $\Teff$ across the stellar solid-angle, which is not the case in the limit of $\tdiff\gg\tcross$. 

\subsection{Bolometric properties}

\begin{figure}
\begin{center}
\includegraphics[width=\columnwidth]{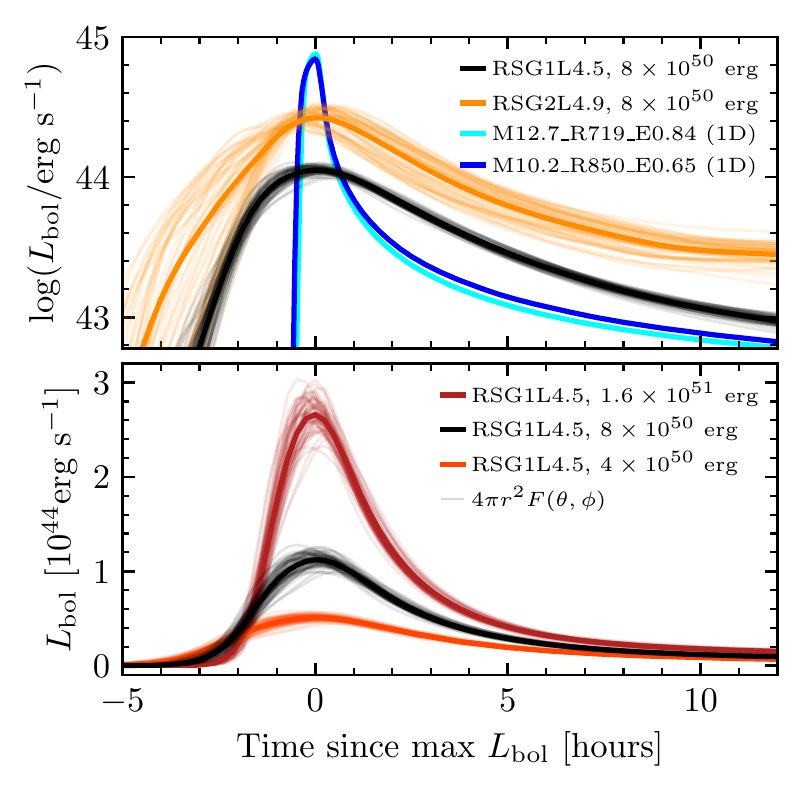}
\caption{Bolometric lightcurves calculated for varied progenitor model (top) and explosion energy (bottom).
Black curves correspond to the fiducial $8\times10^{50}$erg explosion of the RSG1L4.5 model with $\DR=80\Rsun$. Faint lines are calculated from the flux an observer would see at $r=2700\Rsun$ looking back at the star for a particular $\theta,\phi$ location. For comparison, SBO lightcurves from spherically symmetric \MESA+\stella\ explosion models \citep{Goldberg2020} are shown in the upper panel (blues), labeled  M$[\Mej/\Msun]$\_R$[\rphot/\Rsun]$\_E$[\Eexp/10^{51}\mathrm{erg}]$. The x-axis for all 3D curves is the time since the maximum shell-averaged $\Lbol$ (i.e. the peak of the thick curves).}
\label{fig:lightcurves}
\end{center}
\end{figure}

The resulting SBO lightcurves of our 3D explosions are fainter, and longer-duration, than explosions of 1D stellar models. Fig.\ref{fig:lightcurves} shows $\Lbol$ for different explosions, and the explosion properties are summarized in Table~\ref{tab:explosions}.
The upper panel compares $\log(\Lbol)$ of our fiducial explosion (black curves) to a 8$\times10^{50}\mathrm{erg}$ explosion of the RSG2L4.9 model, as well as characteristic 1D explosion models from \citet{Goldberg2020} using \MESA\ and \stella\ \citep{Blinnikov2004,Baklanov2005,Blinnikov2006}
selected for their comparable $\Eexp$, $\Mej$, and radii (blue and cyan).
The 1D lightcurves are corrected for light-travel time as $L_{\mathrm{bol}}(t)=\frac{1}{R/c}\int_{t-R/c}^{t}L\left(t^{\prime}\right)\intdt^{\prime}$ \citep{Shussman2016}, which dominates the SBO duration $t_\mathrm{SBO}\approx R/c$ in 1D explosion models. These 1D explosion lightcurves agree well with the \citet{Shussman2016} semi-analytical models (see also the detailed discussions in \citealt{Kozyreva2020}).
The RSG2L4.9 explosion (orange line in the upper panel) has a smaller $\Mej=3.5\Msun$ in the simulation domain and therefore larger $\vsh$, with similar $\rho_0$ but with greater variety and some material below $10^{-10}\,\mathrm{g\,cm^{-3}}$. The 
greater variety of shock arrival times across the larger $\DR=200\Rsun$ also leads to a greater diversity in velocities and lightcurves along different lines of sight. In both models, the intrinsic $\DR$ dominates over the possible diversity of radii at which SBO would occur at $20<\taus<200$ corresponding to an order of magnitude in $c/\vsh$.
The lower panel compares three different explosions of the RSG1L4.5 model, with $\Eexp=4\times10^{50}\,\mathrm{erg}$, 
$8\times10^{50}\,\mathrm{erg}$, and $1.6\times10^{51}\,\mathrm{erg}$; lower-energy explosions have a longer duration with 
lower peak luminosity compared to higher-energy explosions. In the 3D explosions, the spread of possible SBO signals is 
estimated by $4\pi r^2F_r$ at 64 viewing angles across the surface, shown as faint lines. A full-star simulation would allow
for a more complete sampling of SBO radii which might increase the duration of the lightcurve. This can be quantified with 
explosions at different time snapshots in the star's evolution.

In the 1D model explosions, the intrinsic rise time of the SBO pulse is equal to $\tdiff$ \citep{Nakar2010,Sapir2011}, but the observed duration of the SBO pulse is equal to the light-travel time $R/c$ \citep{Katz2012,Shussman2016,Kozyreva2020}, as $\tdiff<\R/c$ for 1D stellar models with no circumstellar material. In the 3D models, due to the lower densities and larger scale height of material where SBO occurs,
both $\tdiff$ and $\tcross$ dominate $R/c$. Additionally, the diffusion time and the shock traversal time across the inhomogeneous outer imply different scalings with the explosion energy. The shock traversal time, $\tcross\propto\vsh^{-1}\appropto\Eexp^{-1/2}$, whereas {$\tdiff\propto(\rho_0\vsh^2)^{-1}\appropto\Eexp^{-1}$} (this scaling varies for assumptions about the outer density profile, but is steeper than $\Eexp^{-0.8}$; see \citealt{Rabinak2011}, \citealt{Shussman2016}, and others).

\begin{table*}
\begin{center}
\begin{tabular}{| c | c | c | c | c | c | c | c | c | }
\hline Progenitor Model & $\Eexp/10^{51}\mathrm{erg}$ & $\Mej/\Msun$ & $r(\tau_s=100)/\Rsun$ & $L_\mathrm{peak}/10^{44}\mathrm{erg\,s^{-1}}$ &  $\trise$/hr & $\Dthalf$/hr \\ \hline
\hline RSG1L4.5 & $0.4$ & $12.7$ & $\approx820$ & $0.51$ &  $2.33$ & $6.09$ \\ \hline
          & $0.8$*& $12.7$ & $\approx820$ & $1.12$ &  $1.64$ & $4.36$ \\ \hline
         & $1.6$ & $12.7$ & $\approx820$ & $2.66$ &  $1.20$ & $3.02$ \\ \hline
 RSG2L4.9 & $0.8$ & $3.5$ & $\approx960$ & $2.65$ &  $1.56$ & $4.03$ \\ \hline
\end{tabular}
\end{center}
\caption{Summary of the properties of the 3D explosion models, including $\Eexp$ and $\Mej$ in the simulation domain at the time of peak $\Lbol$, the approximate average location of SBO, the peak luminosity, rise time, and duration.
The * denotes the fiducial explosion.}
\label{tab:explosions}
\end{table*}

We define a characteristic SBO duration, $\Dthalf$, as the width of the SBO pulse from half of the peak luminosity on the rise ($\trise$) to half the peak luminosity on the fall ($\tfall$), $\Dthalf=\trise+\tfall$. Characterizing the breakout by $\Dthalf$, rather than $\trise$, is motivated by the fact that when the SBO duration is mediated by the 3D inhomogeneities, the morphology of the $\taus=c/\vsh$ surface will determine the shape of the breakout pulse. This is also evident when comparing the RSG1L4.5 and RSG2L4.9 lightcurves, which have similar $\tcross$ but different morphology where $\taus=c/\vsh$.
For the 1D explosions shown, $\Dthalf=0.55\,\mathrm{hrs}$ for the $850\Rsun$ model and 0.45\,hrs for the $719\Rsun$ model, consistent with $R/c$. 
In the 3D SBOs, $\Dthalf\gg R/c$, and the {relative durations exhibited by the 3D models are consistent with the semi-analytic expectation that}{$\Dthalf\sim\tcross\appropto\DR\ \vsh^{-1}$}, which scales like {$\appropto\Eexp^{-1/2}$ for fixed $\Mej$ and $R$}. {This agreement, rather than steeper dependence on $\Eexp$ expected for $\tdiff>\tcross$ or no dependence on $\Eexp$ if the duration were set by $R/c$, further supports the notion that $\tcross$ is setting the SBO duration for these explosions.}
{In fact, the numerical values of $\Dthalf$ for the explosions summarized in Table~\ref{tab:explosions} agree well with $\Dthalf\approx\tcross$ from Eq.~\ref{eq:tcross}}. 

{In the 3D star, due to the different SBO times at different patches on the stellar surface, individual SBO signals coming from each angular location peak at different times within the span of $\tcross\approx$a few hours, and the duration of each of those individual local breakout signals would be approximately set by $\tdiff\approx$an hour or two as expected for the nearly planar case discussed by, e.g., \cite{Nakar2010} and \citet{Irwin2021}. However, an observer far away from the star does not only see emission coming from a shock breakout along the radial ray directly pointed directly at them. Rather, because emission streaming out of the photosphere at each location on the stellar surface is somewhat isotropic (mediated in part by the stellar surface topography),
and because the difference in light travel time from different patches of the stellar surface ($\approx R/c$) is shorter than $\tcross$,
observers at different viewing locations far away from the star will see a SBO signal that is integrated from the broken-out patches across the portion of the star that they see. This is why the timing of the peak luminosity shown in Fig.~\ref{fig:lightcurves} is more similar for different observer locations, and the duration of the breakout for all observers matches more closely with $\tcross$.}

Since $\tdiff\propto\vsh^{-2}$ but $\tcross\propto\vsh^{-1}$ {for a given star}, the diffusion time may dominate the shock traversal time at lower $\vsh$ (i.e. low $\Eexp$). This occurs when $\vsh<c/(\kappa\rho_0\DR)$, or $\vsh\approx$1,300 $\mathrm{km\,s^{-1}}$ for $\DR=80\Rsun$ and $\rho_0=1.25\times10^{-10}\,\mathrm{g\,cm^{-3}}$. In those cases, {the emission would appear more spherical and} we would also expect less diversity in the observed temperature.

Furthermore, a characteristic luminosity can be predicted {as the internal energy contained outside $\tau<\tausbo$, $E_0$, divided by the relevant timescale, $t_0$, or} $\Lchar=E_0/t_0$. 
Following \citet{Nakar2010}, $E_0\approx4\pi\rshock^2\rho_0\vsh^2(c/\kappa\rho_0\vsh)=4\pi\rshock^2c\vsh/\kappa$, with $\rshock\approx820\Rsun$ for RSG1L4.5 as the average radius of the $\taus\approx100$ surface. 
When $\tcross>\tdiff$, $\Lchar\approx\E_0/\tcross$ {rather than identifying $t_0=\tdiff$}, so
\begin{equation}
    \Lchar\approx4\pi\rshock^2\frac{c}{\kappa}\frac{\vsh^2}{\DR}\appropto\Eexp
    \label{eq:Lchar}
\end{equation}
{where the energy scaling is for fixed stellar properties}. 
{Because $\tcross$ (a few hours) in the 3D models is larger than $\tdiff$ for a 1D stellar model with a barren photosphere (tens of minutes) or $R/c$ ($\approx$half an hour), this entails a SBO signal a factor of $\approx$3-10 times fainter than 1D models predict, as seen in Fig.~\ref{fig:lightcurves}.}
Moreover, the energy scaling is in rough agreement with the peak bolometric luminosity $\Lpeak$ {seen in the 3D explosion models with varied explosion energy}, and the magnitude of $\Lpeak$ matches {Eq.~\ref{eq:Lchar} for} $\Lchar$ within the variance in $\vsh\approx3000-5000\mathrm{km\,s^{-1}}\left(\Eexp/0.8\times10^{51}\mathrm{erg}\right)^{1/2}$ for a given explosion along different lines of sight, which is less than a factor of $2$. 

Finally, when $t>\tcross$ such that most parcels have undergone SBO, but while the fluid elements have not yet doubled their radius, the emission follows the expected $\Lbol\appropto t^{-4/3}$ decline predicted for this planar shock cooling phase \citep[see, e.g.,][]{Nakar2010}. 

The NUV/ULTRASAT band pass (2200-2800\AA) is near the peak of the blackbody for the expected $T=10^{5}-10^{5.5}$K {near the maximum $\Lbol$}, so the rise time and duration of the bolometric luminosity signal are nearly that which would be observed by ULTRASAT. The emission cools into NUV bands beyond the UV/bolometric peak, which does extend the decline in those bands; a full frequency-dependent calculation of SBO emission will yield further predictions about the radiation temperature which would better map to the observables expected from future and current high-energy satellite missions.

\section{Conclusions}

By driving a SN shock through global 3D RHD simulations of RSG envelopes in \Athena, we have explored the effects of the 3D surface on the resulting SBO emission. 
Two important physical differences serve to prolong the SBO duration compared to the spherically symmetric case \citep[see, e.g.][]{Nakar2010,Sapir2011,Katz2012,Sapir2013,Sapir2017}.
First, the intrinsic radiation diffusion time increases due to lower-density material outside the traditional photosphere present in 3D models which contributes non-negligibly to the optical depth near SBO. Most importantly, the bulbous 3D surface with a handful of large-scale plumes spanning $\DR\approx100\Rsun$ yields a variety of shock arrival times at the stellar surface, on a timescale $\tcross\approx\DR/\vsh$. This is an intrinsically 3D phenomenon, setting the observed SBO duration for typical explosion energies. 

Both timescales dominate over the light-travel time across the stellar surface, $R/c$, which is the observed rise time in explosions of 1D stellar progenitors in the absence of circumstellar material \citep[see, e.g.][]{Sapir2013,Shussman2016,Kozyreva2020}. 
This new 3D understanding provides better agreement with the hours-long SBO signal observed in the few existing detections of SN SBO in the UV \citep[e.g.][]{Gezari2008,Schawinski2008,Gezari2010,Gezari2015}. 
The implied longer durations additionally lead to fainter peak luminosities for a given explosion energy, by a factor of $\approx3-10$. Thus, while
useful for constraining the amount of surface asymmetry and the shock velocity
as it reaches the stellar surface, SBO observations cannot independently constrain the stellar radius.

Additionally, at any given point in time near the peak in $\Lbol$, fluid elements across the surface coexist at different stages pre-, mid-, and post-shock-breakout, leading to a diversity in the temperature as a function of optical depth along different lines of sight. Specifics of the radiation spectrum await a multi-group 3D calculation, 
and the 3D nature of the envelope may have further implications for early-time 
spectropolarimetric measurements  \citep[e.g.][]{Leonard2001b,Wang2008,Kumar2016}, 
as well as flash spectroscopy 
\citep[e.g.][]{Khazov2016,Kochanek2019,Soumagnac2020}.

\acknowledgements

We thank the anonymous referee for thoughtful comments which significantly improved the quality of this manuscript. 
We thank Eliot Quataert, William Schultz, Benny Tsang, and Tin Long Sunny Wong for helpful discussions. 

J.A.G. acknowledges NSF GRFP-1650114. This research was 
supported by NSF ACI-1663688 and PHY-1748958, and by NASA
ATP-80NSSC18K0560 and ATP-80NSSC22K0725. 
The Flatiron Institute is supported by the Simons Foundation.
Computational resources were provided by the NASA High-End 
Computing (HEC) program through the NASA Advanced Supercomputing (NAS) 
Division at Ames. We acknowledge support from the Center for Scientific Computing from the CNSI, MRL: an NSF MRSEC (DMR-1720256) and NSF CNS-1725797.
This research made extensive use of the SAO/NASA Astrophysics Data System (ADS).

\bibliographystyle{aasjournal}
\singlespace

\bibliography{RSG3D.bib}



\end{CJK*}
\end{document}